\documentclass[preprintnumbers,superscriptaddress,showpacs,pra]{revtex4}%
\usepackage[colorlinks=true,linkcolor=blue]{hyperref}
\usepackage{amssymb}
\usepackage{amsfonts}
\usepackage{amsmath}
\usepackage{graphicx}%
\providecommand{\U}[1]{\protect\rule{.1in}{.1in}}

\let\stdsection\section
\renewcommand\section{\nopagebreak\stdsection}
\begin{document}
\title{Geometric momentum and angular momentum for charge-monopole system}
\author{S. F. Xiao}
\affiliation{School for Theoretical Physics, School of Physics and Electronics, Hunan
University, Changsha 410082, China}
\affiliation{Department of Physics, Zhanjiang Normal University, Zhanjiang 524048, China}
\author{Q. H. Liu}
\email{quanhuiliu@gmail.com}
\affiliation{School for Theoretical Physics, School of Physics and Electronics, Hunan
University, Changsha 410082, China}
\date{\today}

\begin{abstract}
For a charge-monopole pair, though the definition of the orbital angular
momentum is different from the usual one, and the transverse part of the
momentum that includes the vector potential as an additive term turns out to
be the so-called geometric momentum that is under intensive study recently.
For the charge is constrained on the spherical surface with monopole at the
origin, the commutation relations between all components of geometric momentum
and the orbital angular momentum satisfy the $so(3,1)$ algebra. With
construction of the geometrically infinitesimal displacement operator based on
the geometric momentum, the $so(3,1)$ algebra implies the Aharonov-Bohm phase
shift. The related problems such as charge and flux quantization are also addressed.

\end{abstract}

\maketitle
\preprint{REV\TeX4-1 }

For a particle in the centrally symmetrical potential field, for instance, an
electron interacts with proton that is placed at the origin of the
coordinates, the orbital angular momentum is a conservative quantity, which is
defined by,
\begin{equation}
\mathbf{L}\equiv\mathbf{r\times P},\label{classL}%
\end{equation}
where the position of the particle is defined by its radius vector
$\mathbf{r}$, and $\mathbf{P}$ is the canonical momentum. Once splitting the
momentum $\mathbf{P}$ into a radial $\mathbf{P}^{\parallel}$ and a transverse
part $\mathbf{P}^{\perp}$ in the following, with $r\equiv\left\vert
\mathbf{r}\right\vert $ and $\mathbf{e}_{r}\equiv\mathbf{r/}r$,%
\begin{equation}
\mathbf{P}^{\parallel}\mathbf{\equiv e}_{r}\left(  \mathbf{e}_{r}%
\cdot\mathbf{P}\right)  ,\text{ and \ }\mathbf{P}^{\perp}\mathbf{\equiv
P-e}_{r}\left(  \mathbf{e}_{r}\cdot\mathbf{P}\right)  ,\label{split}%
\end{equation}
we immediately find that in the angular momentum only the transverse part
$\mathbf{P}^{\perp}$ remains and its radial part $\mathbf{P}^{\parallel}$ has
zero effect in it for we have $\mathbf{r}\times\mathbf{P}^{\parallel}=0$ and
an equivalent definition of the orbital angular momentum,
\begin{equation}
\mathbf{L}\equiv\mathbf{r}\times\mathbf{P}=\mathbf{r}\times\mathbf{P}^{\perp}.
\end{equation}
In quantum mechanics, $\mathbf{P}\equiv-i\hbar\nabla$. The splitting of
$\mathbf{P}$ means the gradient operator $\nabla$ for the centrally
symmetrical system is also splittable in the similar fashion. In fact, the
transverse momentum with a given radius $\mathbf{r}$ is the geometric one
which is recently under extensive investigations
\cite{07,liu11,liu13-2,133,liu13-1,135,134,136,137,liu2015,gem,liu17,wang17,iran,liu18}%
, and its explicit form is given shortly, c.f., (\ref{gmx})-(\ref{gmz}). 

In section II, we show how to split the gradient operator $\nabla$ in
spherical polar coordinates. In section III, we apply this splitting to the
charge-monopole system. In final section IV, a brief conclusion is given. In
whole of present paper, only the nonrelativistic motion of the charge is
considered and the spin is ignored.

\section{Radial and transverse decomposition of the gradient operator in
spherical polar coordinates}

For a spherically symmetrical potential field, the most useful coordinates are
the spherical polar ones, and the transformation relation between the
Cartesian coordinates ($x,y,z$) and the spherical polar coordinates
($r,\theta,\varphi$) is,
\begin{equation}
x=r\sin\theta\cos\varphi,\text{ }y=r\sin\theta\sin\varphi,\text{ }%
z=r\cos\theta,
\end{equation}
where $\theta$ is the polar angle from the positive $z$-axis with $0\leq
\theta<\pi$, and $\varphi$ is the azimuthal angle in the $xy$-plane from the
$x$-axis with $0\leq\varphi<2\pi$. The gradient operator in the Cartesian
coordinates, $\nabla_{cart}\equiv\mathbf{e}_{x}\partial_{x}+\mathbf{e}%
_{y}\partial_{y}+\mathbf{e}_{z}\partial_{z}$,\ can be expressed in
($r,\theta,\varphi$) by either of the following two ways,
\begin{subequations}
\begin{align}
\nabla_{sp} &  =\mathbf{e}_{r}\frac{\partial}{\partial r}+\mathbf{e}_{\theta
}\frac{1}{r}\frac{\partial}{\partial\theta}+\mathbf{e}_{\varphi}\frac{1}%
{r\sin\theta}\frac{\partial}{\partial\varphi},\text{ and}\\
\nabla_{sp} &  =\frac{\partial}{\partial r}\mathbf{e}_{r}+\frac{1}{r}%
\frac{\partial}{\partial\theta}\mathbf{e}_{\theta}+\frac{1}{r\sin\theta}%
\frac{\partial}{\partial\varphi}\mathbf{e}_{\varphi}.
\end{align}
These two ways are not identical but equivalent for we have in general the
non-commutation between the derivative with respect to a coordinate and unit
vector of this coordinate in curvilinear coordinates. Thus, we can define the
third one,%
\end{subequations}
\begin{equation}
\nabla_{sp}=\frac{1}{2r}\left(  \frac{\partial}{\partial\theta}\mathbf{e}%
_{\theta}+\frac{1}{\sin\theta}\frac{\partial}{\partial\varphi}\mathbf{e}%
_{\varphi}+\mathbf{e}_{\theta}\frac{\partial}{\partial\theta}+\mathbf{e}%
_{\varphi}\frac{1}{\sin\theta}\frac{\partial}{\partial\varphi}\right)
+\frac{1}{2}\mathbf{e}_{r}\left(  \frac{\mathbf{r}}{r}\cdot\nabla
_{cart}+\nabla_{cart}\cdot\frac{\mathbf{r}}{r}\right)  .
\end{equation}
It is evident that we can define the radial part $\nabla^{\parallel}$ of the
gradient operator in the following,%
\begin{equation}
\nabla^{\parallel}\equiv\frac{1}{2}\left(  \frac{\mathbf{r}}{r}\cdot
\nabla_{cart}+\nabla_{cart}\cdot\frac{\mathbf{r}}{r}\right)  =\mathbf{e}%
_{r}\left(  \frac{\partial}{\partial r}+\frac{1}{r}\right)
\end{equation}
We can define the transverse part $\nabla^{\perp}$ of the gradient operator
$\nabla_{sp}$ in the following,
\begin{align}
\nabla^{\perp} &  =\frac{1}{2r}\left(  \frac{\partial}{\partial\theta
}\mathbf{e}_{\theta}+\frac{1}{\sin\theta}\frac{\partial}{\partial\varphi
}\mathbf{e}_{\varphi}+\mathbf{e}_{\theta}\frac{\partial}{\partial\theta
}+\mathbf{e}_{\varphi}\frac{1}{\sin\theta}\frac{\partial}{\partial\varphi
}\right)  \nonumber\\
&  =\mathbf{e}_{\theta}\frac{1}{r}\frac{\partial}{\partial\theta}%
+\mathbf{e}_{\varphi}\frac{1}{r\sin\theta}\frac{\partial}{\partial\varphi
}-\mathbf{e}_{r}\frac{1}{r}.\label{tran}%
\end{align}
By the "transverse part" we mean that it is perpendicular to the radial
direction $\mathbf{e}_{r}$, which is defined by, with consideration of the
noncommutability between $\mathbf{e}_{r}$ and $\nabla^{\perp}$ \cite{liu2015}%
,
\begin{equation}
\mathbf{e}_{r}\cdot\nabla^{\perp}+\nabla^{\perp}\cdot\mathbf{e}_{r}%
=0.\label{otho}%
\end{equation}
A very puzzling point is: does an explicit inclusion of radial term
$\mathbf{e}_{r}/r$ in the transverse part $\nabla^{\perp}$ mean that
$\nabla^{\perp}$ has component along radial direction? This point can be
clarified in the following: because of relation (\ref{otho}), the radial term
$\mathbf{e}_{r}/r$ is implicitly involved in noncommutability between
$\mathbf{e}_{r}$ and $\mathbf{e}_{\theta}\frac{1}{r}\frac{\partial}%
{\partial\theta}+\mathbf{e}_{\varphi}\frac{1}{r\sin\theta}\frac{\partial
}{\partial\varphi}$, and the implicit and explicit inclusion of the same
radial term $\mathbf{e}_{r}/r$ cancels out in the sense of the relation
(\ref{otho}).

Multiplied $\nabla^{\perp}$ by a factor $-i\hbar$, we obtain the transverse
part of the momentum $\mathbf{P}^{\perp}$ defined by,%
\begin{equation}
\mathbf{P}^{\perp}\mathbf{=}-i\hbar\nabla^{\perp}.\label{geom}%
\end{equation}
In consequence, we have $\mathbf{e}_{r}\cdot\mathbf{P}^{\perp}+\mathbf{P}%
^{\perp}\cdot\mathbf{e}_{r}=0$. The components in Cartesian coordinates give
the standard projections
\cite{07,liu11,liu13-2,133,liu13-1,135,134,136,137,liu2015},
\begin{subequations}
\begin{align}
P_{x}^{\perp} &  =-\frac{i\hbar}{r}(\cos\theta\cos\varphi\frac{\partial
}{\partial\theta}-\frac{\sin\varphi}{\sin\theta}\frac{\partial}{\partial
\varphi}-\sin\theta\cos\varphi),\label{gmx}\\
P_{y}^{\perp} &  =-\frac{i\hbar}{r}(\cos\theta\sin\varphi\frac{\partial
}{\partial\theta}+\frac{\cos\varphi}{\sin\theta}\frac{\partial}{\partial
\varphi}-\sin\theta\sin\varphi),\label{gmy}\\
P_{y}^{\perp} &  =\frac{i\hbar}{r}(\sin\theta\frac{\partial}{\partial\theta
}+\cos\theta),\label{gmz}%
\end{align}
which are three components of the so-called geometric momentum without
presence of the magnetic field. It is in fact two-dimensional spherical
surface case of the following general definition of the geometric momentum
without external field \cite{134,liu17},
\end{subequations}
\begin{equation}
\mathbf{\Pi\equiv-}i\hbar\left(  \nabla_{S}+\frac{M\mathbf{n}}{2}\right)
\rightarrow\mathbf{P}^{\perp}\text{ },\label{PI}%
\end{equation}
in which $\mathbf{\Pi}$ is applicable for a hypersurface in any dimensions
with $M$ denoting the mean curvature of the hypersurface, and for the
two-dimensional spherical surface of a given radius $r$, $\mathbf{\Pi}$ gives
$\mathbf{P}^{\perp}$. However, the geometric momentum including the
contributions of magnetic field in two and three dimensions can be
conveniently defined \cite{liu13-2}, c.f., (\ref{MonopGM}). Here we must
stress that, for a hypersurface, we have a new definition of the gradient
operator,
\begin{equation}
\nabla_{S}\rightarrow\nabla_{S}+\frac{M\mathbf{n}}{2}%
\end{equation}
rather than $\nabla_{S}$ itself, and the new version of the gradient operator
$\nabla_{S}+M\mathbf{n/}2$ has important influences on the propagation of
surface plasmon polaritons on the curved metallic wires \cite{gem}. 

Because of two commuting relations,
\begin{equation}
\lbrack r^{2},\mathbf{L}]=0,\text{ and }[r^{2},\mathbf{\Pi}]=0, \label{comms}%
\end{equation}
we can diagonalize $r^{2}$ and study operators $\mathbf{L}$ and $\mathbf{\Pi}$
for fixed $r$, \cite{CN77} i.e., the charge is constrained on the spherical
surface with the monopole is placed at the origin. The commutation relations
between six operators $[L_{x},L_{y},L_{z},r\Pi_{x},r\Pi_{y},r\Pi_{z}]$ form an
$so(3,1)$ algebra, i.e., algebraic relation (\ref{defL}) together with
following ones are closed in the following ways,
\begin{equation}
\left(  r\mathbf{\Pi}\right)  \mathbf{\times}\left(  r\mathbf{\Pi}\right)
\mathbf{=-}i\hbar\mathbf{L,}\text{and }\mathbf{L\times}\left(  r\mathbf{\Pi
}\right)  =i\hbar\left(  r\mathbf{\Pi}\right)  . \label{so31}%
\end{equation}
In the rest part of the paper, generalization of above analysis to the
charge-monopole system will be made.

\section{Momentum decomposition for charge-monopole system and consequence}

\subsubsection{Angular momentum for charge-monopole system}

A charge-monopole pair can also be treated as a charge in the centrally
symmetrical potential field, but there is an extra angular momentum
originating from a non-trivial field, which was first noted by J.J. Thompson
in 1904 \cite{JJ} and becomes well-known
\cite{CN77,JJ,1931,1944,1969,CN76,1994,2003,2005,2010}. When a static magnetic
monopole with charge $g$ is placed at the origin, the orbital angular momentum
$\mathbf{L}\left(  \mathbf{A}\right)  $ of the electric charge $q$ (for an
electron $q=-e$) can only be defined by,
\begin{equation}
\mathbf{L}\left(  \mathbf{A}\right)  \equiv\mathbf{r\times p-}\eta
\mathbf{e}_{r},\text{ }\eta\equiv\frac{qg}{c},\label{angm}%
\end{equation}
where the momentum $\mathbf{p}$ is the mechanical (or kinematical) one
$\mathbf{p\equiv P-}q\mathbf{A}/c$ with $\mathbf{A}$ being the vector
potential and $\eta\mathbf{e}_{r}$ is the extra angular momentum. It is
readily to verify that this angular momentum satisfies the following $so(3)$
algebra (or the angular momentum algebra),
\begin{equation}
\mathbf{L\left(  \mathbf{A}\right)  \times L}\left(  \mathbf{A}\right)
=i\hbar\mathbf{L\left(  \mathbf{A}\right)  .}\label{defL}%
\end{equation}
When there is no monopole, $\mathbf{A}=0$, $\mathbf{L}(=\mathbf{L}\left(
0\right)  )$ (\ref{classL}) is included as a special case of the
$\mathbf{L\left(  \mathbf{A}\right)  }$ (\ref{angm}).\ Because of the
fundamental importance of the charge-monopole system \cite{dirac}, this
angular momentum is extensively studied in both classical and quantum
mechanics and in various aspects
\cite{CN77,JJ,1931,1944,1969,CN76,1994,2003,2005,2010}. The vector potential
$\mathbf{A}$ can always be chosen to be along the transverse direction because
the magnetic field $\mathbf{B}$ points outward radially,
\begin{equation}
\nabla\times\mathbf{A=B\equiv}\frac{g}{r^{2}}\mathbf{e}_{r}.\label{MagF}%
\end{equation}
The transverse component of the momentum $\mathbf{\mathbf{p}}$ is,
\begin{equation}
\mathbf{\mathbf{p}}^{\parallel}=\mathbf{\Pi}-\frac{q}{c}\mathbf{A\equiv
\Pi(A).}\label{MonopGM}%
\end{equation}
which reduces to the usual geometric momentum (\ref{PI}) $\mathbf{\Pi}\left(
\mathbf{=\Pi(}0\mathbf{)}\right)  $ when the vector potential vanishes
$\mathbf{A}=0$. We have an equivalent definition of the orbital angular
momentum of (\ref{angm}),
\begin{equation}
\mathbf{L(A)}=\mathbf{r\times\Pi(A)-}\eta\mathbf{e}_{r}.
\end{equation}

\subsubsection{Angular momentum and geometric momentum form so(3.1) algebra}

We are going to show that algebraic relations (\ref{so31}) remain with
$\mathbf{\Pi}$ and $\mathbf{L}$ replaced by $\mathbf{\Pi(A)}$ (\ref{MonopGM}%
)\ and $\mathbf{L(A)}$ (\ref{angm}) respectively. To make calculations shorter
and easier, we use an explicit solution \cite{2005} of the vector potential
$\mathbf{A}$ of Eq. (\ref{MagF}),%
\begin{align}
A_{\varphi} &  =\frac{Q}{r}\frac{1-\cos\theta}{\sin\theta},\text{ }\left(
0\leq\theta\leq\frac{\pi}{2}+a\right)  ,\label{AA}\\
A_{\varphi} &  =-\frac{Q}{r}\frac{1+\cos\theta}{\sin\theta},\text{ }\left(
\frac{\pi}{2}-a\leq\theta\leq\pi\right)  ,\label{AB}%
\end{align}
where $A_{\varphi}$ (\ref{AA}) applies to the positive semi-axis $z$, while
$A_{\varphi}$ (\ref{AB}) does to the negative semi-axis $z$, and $a$ is finite
in the range $a\in(0,\pi/2)$, for instance, $a=\pi/4$. In the overlapping area
$\pi/2-a\leq\theta\leq\pi/2+a$, two forms of the vector potential must be
related to each other by a gauge transformation \cite{CN77,2005}. Noting
$\mathbf{e}_{\varphi}=-\mathbf{e}_{x}\sin\varphi+\mathbf{e}_{y}\cos\varphi$,
we have,%
\begin{equation}
\mathbf{A=}A_{\varphi}\mathbf{e}_{\varphi}=-A_{\varphi}\sin\varphi
\mathbf{e}_{x}+A_{\varphi}\cos\varphi\mathbf{e}_{y}=\left(  -A_{\varphi}%
\sin\varphi,A_{\varphi}\cos\varphi,0\right)  .
\end{equation}

Given the vector potential $A_{\varphi}$ (\ref{AA}), three components of the
orbital angular momentum are, respectively,
\begin{subequations}
\begin{align}
L_{x}\left(  \mathbf{A}\right)   &  =y\left(  \Pi_{z}\mathbf{-}\frac{q}%
{c}A_{z}\right)  -z\left(  \Pi_{y}\mathbf{-}\frac{q}{c}A_{y}\right)  -\mu
\sin\theta\cos\varphi,\label{lxmonog}\\
L_{y}\left(  \mathbf{A}\right)   &  =z\left(  \Pi_{x}\mathbf{-}\frac{q}%
{c}A_{x}\right)  -x\left(  \Pi_{z}\mathbf{-}\frac{q}{c}A_{z}\right)  -\mu
\sin\theta\sin\varphi,\label{lymonog}\\
L_{z}\left(  \mathbf{A}\right)   &  =x\left(  \Pi_{y}\mathbf{-}\frac{q}%
{c}A_{y}\right)  -y\left(  \Pi_{x}\mathbf{-}\frac{q}{c}A_{x}\right)  -\mu
\cos\theta. \label{lzmonog}%
\end{align}
Straightforward calculations give two types of the commutation relations,
\end{subequations}
\begin{equation}
\left[  L_{i}\left(  \mathbf{A}\right)  ,\Pi_{j}\left(  \mathbf{A}\right)
\right]  =\left[  L_{i}\left(  \mathbf{A}\right)  ,\left(  \Pi_{j}%
\mathbf{-}\frac{q}{c}A_{j}\right)  \right]  =i\hbar\varepsilon_{ijk}\Pi
_{k}\left(  \mathbf{A}\right)  ,\text{ }%
\end{equation}
and,%
\begin{equation}
\left[  \Pi_{i}\left(  \mathbf{A}\right)  ,\Pi_{j}\left(  \mathbf{A}\right)
\right]  =\left[  \left(  \Pi_{i}\mathbf{-}\frac{q}{c}A_{i}\right)  ,\left(
\Pi_{j}\mathbf{-}\frac{q}{c}A_{j}\right)  \right]  =-i\hbar\varepsilon
_{ijk}\frac{L_{k}\left(  \mathbf{A}\right)  }{r^{2}}.
\end{equation}
Because $r$ acts as a parameter, so the $so(3,1)$ algebra still holds true for
six operators $\left\{  L_{i}\left(  \mathbf{A}\right)  ,\Pi_{i}\left(
\mathbf{A}\right)  \right\}  $ ($i,=1,2,3$).

\subsubsection{Consequence of the $so(3,1)$ algebra}

Let us consider a geometrically infinitesimal displacement operator (GIDO)
\cite{liu18}\ along a small circle around north pole $z=r$, formed by
intersection of the spherical surface and the plane $z=r-\delta z$ with
$\delta z$ is a positive infinitesimal constant. This circle can be
approximated by\textit{\ a small square} in the $xy$-plane and four corners of
the small square are at ($\pm\alpha,\pm\beta$) with $\alpha=\beta\equiv
\sqrt{r\delta z-(\delta z)^{2}/2}\left(  >0\right)  $. The order of the
displacement is anti-clockwise, A($-\alpha/2,-\beta/2$)$\rightarrow$
B($\alpha/2,-\beta/2$)$\rightarrow$ C($\alpha/2,\beta/2$)$\rightarrow$
D($\alpha/2,-\beta/2$)$\rightarrow$ A. We have a GIDO along a small square
$\square$ABCD,
\begin{equation}
G_{\square}\equiv e^{i\frac{\beta\Pi_{y}\mathbf{(A)}}{\hbar}}e^{i\frac
{\alpha\Pi_{x}\mathbf{(A)}}{\hbar}}e^{-i\frac{\beta\Pi_{y}\mathbf{(A)}}{\hbar
}}e^{-i\frac{\alpha\Pi_{x}\mathbf{(A)}}{\hbar}}\approx e^{\frac{\alpha\beta
}{\hbar^{2}}[\Pi_{x}\mathbf{(A)},\Pi_{y}\mathbf{(A)}]}\approx e^{-\frac
{i}{\hbar}\left(  \alpha\beta/r^{2}\right)  L_{z}\left(  \mathbf{A}\right)
}.\label{ro}%
\end{equation}
In calculation, the Baker-Campbell-Hausdorff formula for two possibly
noncommutative operators $X$ and $Y$ as $e^{X}e^{Y}\approx e^{X+Y}e^{[X,Y]/2}$
is used. The operator $L_{z}\left(  \mathbf{A}\right)  $ is explicitly, after
simplification,%
\begin{equation}
L_{z}\left(  \mathbf{A}\right)  =-i\hbar\frac{\partial}{\partial\varphi}%
-\mu.\label{lzA}%
\end{equation}
Thus, the GIDO $G_{\square}$ (\ref{ro}) consists of two factors,
\[
G_{\square}\approx e^{-\frac{i}{\hbar}\left(  \alpha\beta/r^{2}\right)
L_{z}\left(  \mathbf{A}\right)  }=e^{-\frac{i}{\hbar}\left(  \alpha\beta
/r^{2}\right)  (i\hbar\frac{\partial}{\partial\varphi})}e^{i\frac{\mu}{\hbar
}\left(  \alpha\beta/r^{2}\right)  }%
\]
in which $e^{-\frac{i}{\hbar}\left(  \alpha\beta/r^{2}\right)  (i\hbar
\frac{\partial}{\partial\varphi})}$ is a real rotational operator in the
$xy$-plane, and $e^{i\frac{\mu}{\hbar}\left(  \alpha\beta/r^{2}\right)  }$
stands for a phase factor. The angle of the rotation is the solid angle
element $\delta\Omega$ corresponding to the fixed spherical cap of area
element $\delta S\equiv r^{2}\delta\Omega$ with the circle as its boundary,
\begin{equation}
\delta\Omega\approx\frac{\alpha\beta}{r^{2}},
\end{equation}
and the amount of the phase shift is,
\begin{equation}
\frac{\mu}{\hbar}\delta\Omega.\label{phaseshift}%
\end{equation}
It is in fact the Aharonov-Bohm phase for it can be made manifestly,
\begin{equation}
\frac{\mu}{\hbar}\delta\Omega\approx\frac{q}{\hbar c}\left(  \frac{g}{r^{2}%
}\delta S\right)  =\frac{q}{\hbar c}\delta\phi,\label{ABphase}%
\end{equation}
where $\delta\phi\equiv g\delta S/r^{2}$ is the magnetic flux through the area
element $\delta S$. Next, we will show that the flux $\delta\phi$ is quantized.

Given the vector potential $A_{\varphi}$ (\ref{AB}), the same manner of
calculations gives the same conclusions. However, there is an interesting
difference for the operator $L_{z}$ which is now given by,%
\begin{equation}
L_{z}\left(  \mathbf{A}\right)  =-i\hbar\frac{\partial}{\partial\varphi}+\mu.
\label{lzB}%
\end{equation}
Two forms (\ref{lzA}) and (\ref{lzB}) of $L_{z}\left(  \mathbf{A}\right)  $
must give identical set of the energy eigenvalues, i.e., $L_{z}\left(
\mathbf{A}\right)  =m\hbar$ ($m=0,\pm1,\pm2,...$). It leads to the well-known
quantization of $\mu$ whose allowable values are given by,
\begin{equation}
\mu=\frac{1}{2}n\hbar,\text{ }(n=0,\pm1,\pm2,...).
\end{equation}
It is the right condition imposing upon simultaneous quantization of the
magnetic and electric charge with $q=-e$, firstly obtained by Dirac
\cite{dirac}. Substituting this result into (\ref{phaseshift}), we get the
quantization of the phase shift for a given small circle of the solid angle
element $\delta\Omega,$
\begin{equation}
\frac{\mu}{\hbar}\delta\Omega=\frac{1}{2}m\delta\Omega. \label{last2}%
\end{equation}
It amounts to the flux quantization for we have from combination of results
(\ref{ABphase}) and this one (\ref{last2}),
\begin{equation}
\frac{q}{\hbar c}\delta\phi=\frac{1}{2}m\delta\Omega,\text{ i.e., }\delta
\phi=m\phi_{0}\frac{\delta\Omega}{4\pi}, \label{last}%
\end{equation}
where $\phi_{0}\equiv hc/q\ $is the basic magnetic flux quantum with charge
$q$, agreement with the superconducting magnetic flux quantum where the charge
$q=-2e$ for the Cooper pairs. This result (\ref{last}) also implies that the
total magnetic flux of emitting from a monopole is quantized as $\phi
=m\phi_{0}$ with $\delta\Omega\rightarrow4\pi$.

Above results from (\ref{ro}) to (\ref{last}) are applicable to any point on
the spherical surface, independent from the north pole we start from, because
any point can always be taken as the north pole due to the spherical symmetry.

\section{Conclusions}

The gradient operator in spherical polar coordinates can be decomposed into
the radial and transverse parts. With this decomposition, we show that for the
electron-monopole system, the commutation relations between all components of
the geometric momentum and the orbital angular momentum form an $so(3.1)$
algebra. When there is no monopole but the particle is also constrained on the
spherical surface, the algebraic relations remain which is the special case of
present results with magnetic charge $g=0$. With construction of the GIDO
based on the geometric momentum, we demonstrate that as action of the GIDO on
a quantum state along a small circle around any point on the spherical
surface, the geometric momentum results in a real rotation and a phase shift,
where the angle of the rotation is the solid angle element corresponding to
the small spherical cap with the circle as its boundary, and the phase shift
is the Aharonov-Bohm phase shift which is related to the flux quantization.

\begin{acknowledgments}
This work is financially supported by National Natural Science Foundation of
China under Grant No. 11675051.
\end{acknowledgments}

\end{document}